\documentclass[12pt]{article}

\def\be{\begin{equation}}
\def\ee{\end{equation}}
\def\bea{\begin{eqnarray}}
\def\eea{\end{eqnarray}}

\usepackage{amssymb,latexsym}
\usepackage{graphicx}

\def\lesssim{\mathrel{\mathpalette\fun <}}
\def\gtrsim{\mathrel{\mathpalette\fun >}}
\def\fun#1#2{\lower3.6pt\vbox{\baselineskip0pt\lineskip.9pt
  \ialign{$\mathsurround=0pt#1\hfil##\hfil$\crcr#2\crcr\sim\crcr}}}
\input epsf
 
 \usepackage{epsfig}
\usepackage[section]{placeins}

\makeatletter \@addtoreset{equation}{section} \makeatother

\begin{document}
\begin{titlepage}
\thispagestyle{empty}
\begin{flushright}
    SU-ITP-2007-4\\
April 4, 2007
\end{flushright}

\vspace{30pt}

\begin{center}
    { \LARGE{\bf Testing String Theory with CMB}}

    \vspace{40pt}

  {\large
    Renata Kallosh and Andrei Linde}

    \vspace{10pt}

    \vspace{10pt} { \ Department of Physics,
    Stanford University, Stanford, CA 94305}

    \vspace{20pt}
 \end{center}
 
 \begin{abstract}

Future detection/non-detection of tensor modes from inflation in CMB observations presents a  unique way to test certain features of string theory. Current limit on the ratio of tensor to scalar perturbations, $r=T/S$, is $r\lesssim 0.3$, future detection may take place for $r \gtrsim 10^{-2}-10^{-3}$. At present all known string theory inflation models predict tensor modes well below the level of detection. Therefore a possible experimental discovery of tensor modes may present a challenge to string cosmology.  

The strongest bound on $r$ in string inflation follows from the observation that in most of the  models based on the KKLT construction, the value of the Hubble constant $H$ during inflation must be smaller than the gravitino mass. For the gravitino mass in the usual range, $m_{3/2} \lesssim {\cal O}(1)$ TeV, this  leads to an extremely strong  bound $r \lesssim 10^{{-24}}$. A discovery of tensor perturbations with $r\gtrsim 10^{-3}$ would imply that the gravitinos in this class of models are superheavy,  $m_{3/2}\gtrsim 10^{13}$ GeV. This would have important implications for particle phenomenology based on string theory.

\end{abstract}

\end{titlepage}

\section{Introduction}

Cosmological observations may provide several interesting ways of testing string theory, which is important for its further development. For example, 
a discovery of the cosmological acceleration corresponding to the existence of the cosmological constant $\Lambda \sim 10^{-120}$ (in Planck units $M_{p} = 1$, where $M_{p} = 2.435\times 10^{18}$ GeV)  initially was viewed  as a problem for string theory. For a while is was not  known how to  describe an accelerating 4D universe in a  vacuum state with a positive energy density. Eventually the problem was resolved by the  KKLT construction \cite{KKLT} (developing on  \cite{GKP}), which allowed to explain acceleration in a metastable vacuum state. Earlier and further investigation of these issue \cite{Douglas}, combined with the ideas of eternal inflation \cite{Vilenkin:xq,Eternal}, resulted in the development of the idea of inflationary multiverse \cite{Eternal,book} and string landscape scenario \cite{Susskind:2003kw}, which may have important implications for the general methodology of theoretical physics.

There are some other ways in which cosmology can be used for testing string theory. Much attention of string theory  and cosmology communities during the recent few years, starting with \cite{Copeland:2003bj}, was dedicated to the possible future detection of cosmic strings produced after inflation \cite{Kachru:2003sx,HenryTye:2006uv}. It is viewed as a possible window of a string theory into the real world. If detected, cosmic strings in the sky may  test  various ideas in string theory and cosmology. 

One may also try to check which versions of string theory lead to the best description of inflation, in agreement with the existing measurements of the anisotropy of the cosmic microwave background radiation produced by  scalar perturbations of metric \cite{Mukh}. These measurements provide an important information  about the structure of the inflaton potential \cite{LL,WMAP,Tegmark,Kuo:2006ya}. 
In particular, observational constraints on the amplitude of scalar perturbations, in the slow roll approximation, imply that
\begin{eqnarray}
 \label{eq:V} {V^{{3/2}}\over V'} \simeq 5\times10^{-4} \ ,
\end{eqnarray}
whereas the spectral index of the scalar perturbations  is given by
\begin{eqnarray}
 \label{eq:ns} n_{s}  = 1 - 3 \left({V'\over V}\right)^{2}+ 2   {V''\over V} \approx 0.95 \pm 0.02 \ 
\end{eqnarray}
if the  ratio of tensor perturbations to scalar perturbations is sufficiently small, $r\ll 0.1$. For larger values of $r$, e.g. for $r\sim 0.2$, $n_s= 0.98\pm 0.02$.

However, these data give rather indirect information about $V$: One can reduce the overall scale of energy density by many orders of magnitude, change its shape, and still obtain scalar perturbations with the same properties. 

In this sense, a  measurement of the tensor perturbations (gravitational waves) \cite{Star}, or of the tensor-to scalar ratio $r=T/S$, would be especially informative, since it is directly related to the value of the inflationary potential and the Hubble constant during inflation \cite{LL},
\be\label{eq:rvh}
r = 8 \left(\frac{V'}{V}\right)^2 \approx 3\times  10^{7}~V  \sim 10^{8}~ H^{2}.
\ee
The last part of this equation follows from Eg. (\ref{eq:V}) and from the Einstein equation $H^{2} = V/3$.

The purpose of this note is to address the issues of string cosmology in view of the possibility that tensor modes in primordial spectrum may be detected.  
We will argue here that the possible detection of tensor modes from inflation may have dramatic consequences for string theory and for fundamental physics in general. The current limit on the ratio of tensor to scalar fluctuations is $r<0.3$. During the next few years one expects to probe tensor modes with  $r\sim 0.1$ and gradually reach the level of $r\sim 0.01$. It is believed that probing below $r\sim 10^{-2}- 10^{-3}$ will be ``formidably difficult'' \cite{Peacock:2006kj}.
However, the interval between $r=0.3$ and $r \sim 10^{-3}$ is quite large, and it can be probed by the cosmological observations.  

Expected amplitude of tensor perturbations in stringy inflation appears to be very low, $r\ll 10^{-3}$, see in particular \cite{Baumann:2006cd,Bean:2007hc}. In Section 2 we will briefly review their results, as well as  some other recent results concerning string theory inflation \cite{Kallosh:2007ig}. In Section 3 we give some independent arguments using the relation between the maximal value of the Hubble constant during inflation  and the gravitino mass \cite{Kallosh:2004yh}, which suggest that in the  superstring models based on generic KKLT construction  the amplitude of tensor perturbations in string theory inflation with $m_{3/2} \lesssim 1$ TeV should be extremely small, $r \lesssim 10^{-24}$. 

One could argue therefore that the experimental detection of tensor modes would be in a contradiction with the existing models of string cosmology. Let us remember, however, that many of us did not expect the discovery of the tiny cosmological constant $\Lambda \sim 10^{-120}$, and that it took some time before we learned how to describe acceleration of the universe in the context of string theory. Since there exists a class of rather simple non-stringy inflationary models predicting $r$ in the  interval $0.3 \lesssim r \lesssim 10^{-3}$ \cite{Linde:1983gd,deVega:2006un,lindeparis,Destri:2007pv,Freese:1990rb,Savage:2006tr},  it makes a lot of sense to look for tensor perturbations using the CMB experiments.  It is important to think, therefore, what will happen if the cosmological observations will discover tensor perturbations in the range $10^{{-3}}< r<0.3$. As we will see, this result would not necessarily contradict string theory, but it may have important implications for the models of string theory inflation, as well as for particle phenomenology based on string theory.

\

\section{Tensor modes in the simplest inflationary models}

Before discussing the amplitude of tensor modes in string theory, we will briefly mention what happens in general non-stringy inflationary models.

The predicted value of $r$ depends on the exact number of e-foldings $N$ which happened after the time when the structure was formed on the scale of the present horizon. This number, in turn, depends on the mechanism of reheating and other details of the post-inflationary evolution. 

For $N \sim 60$, one should have $r \sim 0.14$ for the  simplest chaotic inflation model $m^{2}\phi^{2}/2$, and $r \sim 0.28$ for the  model $\lambda\phi^{4}/4$. In the  slow-roll approximation, one would have $r = 8/N$ for the model $m^{2}\phi^{2}/2$ and $16/N$ for the model $\lambda\phi^{4}/4$ \cite{LL}. 

If one considers the standard spontaneous symmetry breaking model with the potential
\be
V = -{m^2\over 2}\phi^2 +{\lambda\over 4}\phi^4+{m^4\over 4 \lambda} = {\lambda\over 4}(\phi^{2} -v^{2})^{2} \ ,
\ee
with $v = m/\sqrt\lambda$, it leads to chaotic inflation with the tensor to scalar ratio which can take any value in the interval $10^{-2} \lesssim r \lesssim 0.3$, for $N \sim 60$. The value of $r$ depends on the scale of the spontaneous symmetry breaking $v$  \cite{deVega:2006un,lindeparis}, see Fig.~1. The situation in the so-called natural inflation model \cite{Freese:1990rb} is very similar   \cite{Savage:2006tr}, except for the upper branch of the curve above the green star (the first star from below) shown in Fig. 1, which does not appear in natural inflation.

\begin{figure}[h!]
\centering\leavevmode\epsfysize=4.5cm \epsfbox{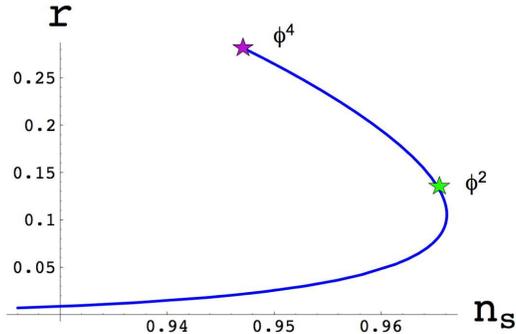}  
\caption{Possible values of $r$ and $n_{s}$ in the theory ${\lambda\over 4}(\phi^{2}-v^{2})^{2}$ for different initial conditions and different $v$, for $N = 60$.  In the small $v$ limit, the model has the same predictions as the theory $\lambda\phi^{4}/4$. In the large $v$ limit it has the same predictions as the theory $m^{2}\phi^{2}$. The branch above the green star (the first star from below) corresponds to inflation which occurs while the field rolls down from  large $\phi$, as in the simplest models of chaotic inflation. The lower branch corresponds to the motion from $\phi = 0$, as in new inflation. }
\label{cmbquadr}
\end{figure}

If one considers chaotic inflation with the potential including terms $\phi^{2}$,  $\phi^{3}$ and  $\phi^{4}$, one can considerably alter the properties of inflationary perturbations \cite{Hodges:1989dw} and cover almost all parts of the area in the $(r,n_{s})$ plane allowed by the latest observational data 
\cite{Destri:2007pv}.  

However, in all of these models the value of $r$ is large because the change of the inflation field during the last 60 e-folds of inflation is greater than $M_{p}=1$ \cite{Lyth:1996im}, which is not the case in many other inflationary models, such as new inflation \cite{new} and hybrid inflation \cite{Hybrid},  see \cite{Lyth:1996im,Efstathiou:2006ak} for a discussion of this issue. Therefore the bet for the possibility of the observational discovery of tensor modes in non-stringy inflationary models would be a bet for the triumph of simplicity over majority.

\section{Existing  models of string theory inflation do not predict a detectable level of tensor modes}

String theory at present has produced two classes of models of inflation: brane inflation and modular inflation, see \cite{HenryTye:2006uv,Kallosh:2007ig,Cline:2006hu} for recent reviews.  The possibility of a significant level of tensor modes in known brane inflation models was carefully investigated by several authors. The following conclusion has been  drawn from our analysis of the work performed by Bean, Shandera, Tye, and Xu \cite{Bean:2007hc}. They compared the  brane inflationary model to recent cosmological data, including WMAP 3-year cosmic microwave background (CMB) results, Sloan Digital Sky Survey luminous red galaxies (SDSS LRG) power spectrum data and Supernovae Legacy Survey (SNLS) Type 1a supernovae distance measures. When they 
used the bound on the distance in the warped throat geonetry derived by Baumann and McAllister 
\cite{Baumann:2006cd}, it became clear that in all currently known models of brane inflation (including DBI models \cite{Alishahiha:2004eh}) the resulting primordial spectrum could not simultaneously have significant deviation from the slow roll behavior and satisfy  the bound \cite{Baumann:2006cd}. Moreover the slow roll inflation models that satisfy the bound have very low tensors not measurable by current or even upcoming experiments. The known models of brane inflation include the motion of a D3 brane  down a single throat in the framework of the KKLMMT scenario \cite{Kachru:2003sx}. In short, the bound on an inflaton  field, which is interpreted as a distance between branes, does not permit fields with vev's of Planckian scale or larger, which would lead to tensor modes. A work on the improved derivation of the bound including the breathing mode of the internal geometry is in progress \cite{DL}.

At present, there is still a hope that it may be possible  to go beyond the simplest models of brane inflation  and evade the constraint on the field range. 
However, this still has to be done before one can claim that string theory has a reliable class of brane inflation models predicting tensor modes, or, on the contrary, that brane inflation predicts a non-detectable level of tensor modes.

All known models of modular inflation in string theory (no branes) do not predict a detectable level of gravity waves \cite{Cline:2006hu}, \cite{Kallosh:2007ig}. The only string theory inspired version  of assisted inflation model \cite{Liddle:1998jc}, N-flation \cite{Dimopoulos:2005ac}, would predict a significant level of tensors, as in chaotic and natural inflation \cite{Linde:1983gd,Freese:1990rb,Savage:2006tr}, if some assumptions underlying the derivation of this model would be realized.
The main assumption  is that in the effective supergravity model with  numerous complex moduli, 
$
t_n= {\phi_n\over f_n}+i M^2 R^2_n,
$
all moduli $R_n^2$ quickly go to their minima. Then only the axions ${\phi_n\over f_n}$ remain to drive inflation.  The reason for this assumption is that the K\"ahler potential depends only on the volume modulus of all two-cycles, $R_n^2 =-{i\over  2M^2} (t_n-\bar t_n)$, but is does not depend on the axions ${\phi_n\over f_n}= {1\over 2}( t_n+\bar t_n)$, so one could expect that the axion directions in the first approximation remain flat. Recently this issue was re-examined in \cite{Kallosh:2007ig}, and it was found that in all presently available models this assumption is not satisfied. The search for models in various regions of the  string theory landscape which would support  assumptions  of N-flation  is  in progress \cite{NM}.

Thus at present we are unaware of any string inflation models predicting the detectable level of gravitational waves. However, a search for such models continues. We should mention here possible generalizations on N-flation, new types of brane inflation listed in Sec. 5 of \cite{Bean:2007hc} and some work in progress on DBI models in a more general setting \cite{Eva}.

We may also try to find a string theory generalization of  a class of  inflationary models in $N=1$ $d=4$ supergravity, which has shift symmetry and predict large tensor modes. One model is a supergravity version \cite{Kawasaki:2000yn} of chaotic inflation, describing fields $\Phi$ and $X$ with
\be
K= {1\over 2} (\Phi+\bar \Phi)^2+X\bar X \ ,  \qquad W= m\Phi X \ .
\label{chaotic}\ee
This model effectively reproduces the simplest version of chaotic inflation with $V={1\over 2} m^2\phi^2$,  where the inflaton field is $\phi=i(\Phi-\bar \Phi)$. Here the prediction for $r$, depending on the number of e-foldings, is $0.14\lesssim r \lesssim 0.20$.

Another model is a supergravity version \cite{Kallosh:2007ig} of natural inflation \cite{Freese:1990rb}. 
\be
K={1\over 2} (\Phi+\bar \Phi)^2\ ,  \qquad  W= w_0 + B e^{-b \Phi}\ .
\label{axionvalley}\ee
This model has an axion valley potential in which the radial part of the complex field quickly reaches the minimum. Therefore this model effectively reproduces  natural  inflation with the axion playing the role of the inflaton with potential $V=V_0(1 -\cos(b\phi))$ where $\phi=i(\Phi-\bar \Phi)$. Here the possible range of  $r$, depending on the number of e-foldings and the axion decay constant $(\sqrt 2 \,b)^{-1}$, is approximately  $5\times 10^{-3} \lesssim r\lesssim 0.20$ \cite{Savage:2006tr}.

Both models have one feature in common. They require  shift symmetry of the canonical K\"ahler potential $K={1\over 2} (\Phi+\bar \Phi)^2$,
\be
\Phi  \rightarrow \Phi + i \delta\ ,  \qquad  \delta=\bar \delta \ .
\ee
The inflaton potential appears because this shift symmetry is slightly broken by the superpotential.

If supersymmetry will be discovered in  future, 
one would expect that inflationary potential should be represented by a supergravity potential, or even better,   by the supergravity effective potential derivable from string theory. It is gratifying that at least some supergravity models capable of prediction of large amplitude of tensor perturbations from inflation are available.

So far, neither of the supergravity models in (\ref{chaotic}), (\ref{axionvalley}) with detectable level of gravity waves was derived from string theory.\footnote{There is a difference between arbitrary $N=1$, $d=4$  supergravity model of the general type and models derived from string theory where various fields in effective supergravity theory have some higher-dimensional interpretation, like volumes of cycles, distance between branes etc. However, 
there are situations in string theory when the actual value of the K\"ahler potential is not known and therefore models like (\ref{chaotic}), (\ref{axionvalley}) are not {\it a priori} excluded.}  
It would be most important to study all possible corners of the landscape in a search of models which may eventually predict  detectable tensor fluctuations, or prove that it is not possible. 
The future data on $r$ will make a final judgment on the theories discussed above.

If some models in string cosmology  with $r>10^{-3}$ will be found, one can use the detection of gravity waves for testing  models of moduli stabilization in string theories, and in this way relate cosmology to particle physics. The main point here is that the value of the Hubble constant during inflation is directly measurable in case that gravity waves are detected.

\section{Scale of SUSY breaking, the gravitino mass, and the amplitude of the gravitational waves in string theory inflation}

So far, we did not discuss relation of the new class of models with particle phenomenology. This relation is rather unexpected and may impose strong constraints on particle phenomenology and on inflationary models: In simplest models based on the KKLT mechanism the Hubble constant $H$  should be  smaller than the present value of the gravitino mass  \cite{Kallosh:2004yh},
\begin{equation}\label{constr}
 H \lesssim m_{{3/2}} \ .
\end{equation}
The reason for this bound is that the mass of gravitino at the supersymmetric KKLT minimum with $DW=0$ before the uplifting is given by 
$
3m_{{3/2}}^2= |V_{AdS}| 
$.
Uplifting of the AdS minimum to the present nearly Minkowski vacuum is achieved by adding to the potential a term of the type of $C/\sigma^{n}$, where $\sigma$ is the volume modulus and $n=3$ for generic compactification and $n=2$ for the highly warped throat geometry. Since the uplifting is less significant at large $\sigma$, the barrier created by the uplifting  generically is a bit smaller than $|V_{AdS}|$. Adding the energy of the inflaton field leads to an additional uplifting. Since it is also proportional to an inverse power of the volume modulus, it is greater at the minimum of the KKLT potential than at the top of the barrier. Therefore adding a large vacuum energy density  to the KKLT potential, which is required for inflation, may uplift the minimum to the height greater than the height of the barrier, and  destabilize it, see Fig. 2. This leads to the bound (\ref{constr}). 

\begin{figure}[h!]
\centering\leavevmode\epsfysize=5.5cm \epsfbox{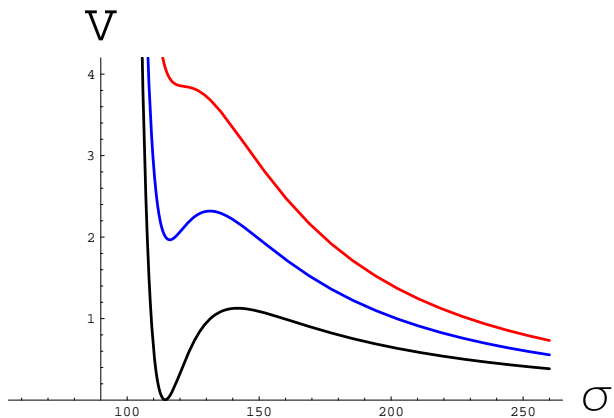} \caption[fig2]
{The lowest curve with dS minimum is the potential of the KKLT model. The second one shows what happens to the volume modulus potential when the inflaton potential $V_{\rm infl}={V(\phi)\over \sigma^3}$ added to the KKLT potential. The top curve shows that when the  inflaton potential becomes too large, the barrier disappears, and the internal space decompactifies. This explains the constraint $H\lesssim   m_{3/2}$.  } \label{2}
\end{figure}

One should note that an exact  form of this bound  is a bit more complicated than (\ref{constr}), containing additional factors which depend logarithmically on certain parameters of the KKLT potential. However,  unless these parameters are exponentially large or exponentially small, one can use the simple form of this bound, $H \lesssim m_{{3/2}}$.

Therefore if one believes in the standard SUSY phenomenology with $m_{{3/2}} \lesssim O(1)$ TeV, one should find a realistic particle physics model where  the nonperturbative string theory dynamics occurs at the LHC scale (the mass of the volume modulus is  not much greater than the gravitino mass), and inflation occurs  at a density at least 30 orders of magnitude below the Planck energy density. Such models are possible, but their parameters should be substantially different from the parameters used in all presently existing models of string theory inflation.

An interesting observational consequence of this result is that the amplitude of the gravitational waves in all string inflation models of this type should be extremely small. Indeed, according to Eq. (\ref{eq:rvh}), one has 
$ {r} \approx 3\times  10^{7}~V  \approx 10^{8}~ H^{2}$, which implies that 
\begin{equation}\label{bound}
r \lesssim 10^{8}~m_{{3/2}}^{2}  \ ,
\end{equation}
in Planck units. In particular, for $m_{{3/2}} \lesssim 1$ TeV $\sim 4 \times 10^{-16}~ M_{p}$, which is in the range most often discussed by SUSY phenomenology, one has
\begin{equation}
r \lesssim 10^{-24} \ .
\end{equation}
If CMB experiments find that $r \gtrsim 10^{-2}$, then this will imply, in the class of theories described above, that 
\begin{equation}
m_{{3/2}} \gtrsim 10^{-5}~ M_{p} \sim 2.4 \times 10^{13}~{\rm GeV}  \ ,
\end{equation}
which is 10 orders of magnitude greater than the standard gravitino mass range discussed by particle phenomenologists.

There are several different ways to address this problem. First of all, one may  consider KKLT models with the racetrack superpotential containing at least two exponents and find such parameters that the supersymmetric minimum of the potential even before the uplifting  occurs at zero energy density   \cite{Kallosh:2004yh}, which would mean $m_{3/2} = 0$. Then, by a slight change of parameters one can get the gravitino mass squared much smaller than the height of the barrier, which removes the constraint $H \lesssim m_{{3/2}}$.

If we want to increase the upper bound on $H$ from $1$ TeV up to $10^{13}$ GeV  for  $m_{{3/2}} \sim 1$ TeV, we would need to fine-tune the parameters of the model of Ref.  \cite{Kallosh:2004yh} with a very high accuracy. Therefore it does not seem easy to increase the measurable value of $r$ in the model of  \cite{Kallosh:2004yh} from $10^{-24}$ up to $10^{-3}$. However,  this issue requires a more detailed analysis, since this model is rather special: In its limiting form, it describes a supersymmetric Minkowski vacuum without any need of uplifting, and it has certain advantages with respect to vacuum stability being protected by supersymmetry were discussed in \cite{Blanco-Pillado:2005fn}. Therefore it might happen that this model occupies a special place in the landscape which allows a natural way towards large $r$.

We will discuss now several other models of moduli stabilization  in string theory to see whether one can overcome  the  bound (\ref{bound}).

A new class of moduli stabilization in M-theory was recently developed in \cite{Acharya:2007rc}.   In particular cases studied numerically, the height of the barrier after the uplifting is about $V_{barrier}\approx 50~m_{{3/2}}^{2}$, in some other cases,  $V_{barrier}\leq {\cal O}(500) \ m_{{3/2}}^{2}$ \cite{private}. It seems plausible that for this class of models, just as in the simplest KKLT models,   the condition that $V_{barrier}\geq 3H^2$ is required for stabilization of moduli during inflation. Since the gravitino mass in this model is in the range from 1 TeV to 100 TeV, the amplitude of the tensor modes is expected to be negligibly small.

Another possibility is to consider the large volume compactification models with stringy $\alpha'$ corrections taken into account \cite{Balasubramanian:2005zx}. At first glance, this also does not seem to help.  The AdS minimum at which moduli are stabilized before the uplifting is not supersymmetric, which means that generically in AdS minimum
$
3m_{{3/2}}^2= |V|_{AdS} + e^{K}|DW|^2\geq |V|_{AdS} 
$.
Upon uplifting, generically the height of the barrier is not much different from the absolute  value of the potential in the AdS minimum,
$
V_{barrier} \sim |V|_{AdS}
$.
As the result, the situation with the destabilization during inflation may seem even more difficult than in the simplest KKLT models: the extra term due to broken supersymmetry $e^{K}|DW|^2\neq 0$ tends to increase the gravitino mass squared as compared to $|V|_{AdS}$. This decreases the ratio of  the height of the barrier after the uplifting to the gravitino mass squared. However, a more detailed investigation of this model is required to verify this conjecture. As we already mentioned, an important assumption in the derivation of the constraint $H \lesssim m_{3/2}$ in the simplest version of the KKLT model is the absence of exponentially large parameters. Meanwhile the volume of compactification in \cite{Balasubramanian:2005zx} is exponentially large. One should check whether this can help to keep the vacuum stabilized for large $H$.

But this class of models offers another possible way to address the low-H problem: In the phenomenological models based on   \cite{Balasubramanian:2005zx} the gravitino mass can be extremely large.
Phenomenological models with superheavy gravitinos were also considered in   \cite{DeWolfe:2002nn,Arkani-Hamed:2004fb}. In particular, certain versions of the split supersymmetry models allow gravitino masses in the range of $10^{13} - 10^{14}~{\rm GeV}$ \cite{Arkani-Hamed:2004fb}. Therefore in such models the constraint $H \lesssim m_{3/2}$ is quite consistent with the possibility of the discovery of tensor modes with $ 10^{-3}\lesssim r \lesssim 0.3$ if the problems with constructing the corresponding inflationary models discussed in the previous section will be resolved.

\vskip 0.2cm

We would like to stress that we presented here only a first scan of possibilities available in string cosmology with regard to detectability of the tensor modes, and so far the result is negative. More studies are required to have a better prediction of $r$ in string cosmology. It would be most important  either to  construct a reliable inflationary model in string theory predicting tensors with $ 10^{-3}\lesssim r \lesssim 0.3$, or prove a no-go theorem.  If tensor modes will not be detected, this issue will disappear; the attention will move to more precise values of the tilt of the spectrum $n_s$, non-gaussianity, cosmic strings and other issues which will be clarified by  observations in the next few years. 

However, a possible discovery of tensor modes may force us to reconsider several basic assumptions of string cosmology and particle phenomenology. In particular, it may imply that the gravitino must be superheavy.  Thus, investigation of gravitational waves produced during inflation may serve as a unique source of information about string theory and about the fundamental physics in general.

\

\noindent{\large{\bf Acknowledgments}}

\noindent  We are grateful to D. Baumann, R. Bean, S.E. Church, G. Efstathiou, S. Kachru, L. Kofman, D. Lyth, L. McAllister,  V. Mukhanov,  S. Shenker,  E. Silverstein and H. Tye   for very stimulating discussions.   This work  was supported by NSF
grant PHY-0244728.

\end{document}